\documentclass[useAMS,usenatbib]{mn2e}
\usepackage{epsfig}

\title[Clear sky fraction above Indonesia]{Clear sky fraction above Indonesia:
an analysis for astronomical site selection}
\author[T. Hidayat et al.]{T. Hidayat,$^{1}$\thanks{E-mail:
taufiq@as.itb.ac.id}, P. Mahasena,$^{1}$
B. Dermawan,$^{1}$
T. W. Hadi,$^{2}$ 
P. W. Premadi$^{1}$ 
\newauthor
and D. Herdiwijaya$^{1}$
\\
$^{1}$Bosscha Observatory and Astronomy Research Division, FMIPA, Institut Teknologi Bandung\\
$^{2}$Atmospheric Science Research Division, FITB, Institut Teknologi Bandung,
Jl. Ganesha 10 Bandung 40132, Indonesia}
\begin{document}

\date{Accepted 2012 August 26. Received 2012 August 17; in original form 2012 May 6}

\pagerange{\pageref{firstpage}--\pageref{lastpage}} \pubyear{2012}

\maketitle

\label{firstpage}

\begin{abstract}
We report a study of cloud cover over Indonesia based on meteorological satellite data,
spanning over the past 15 years (from 1996 to 2010) in order to be able to select a new
astronomical site capable to host a multi-wavelength astronomical observatory.  
High spatial resolution of meteorological satellite data acquired from {\it Geostationary Meteorological Satellite 5} ({\it GMS 5}), {\it Geostationary Operational Environmental Satellite 9} ({\it GOES 9}), and {\it Multi-functional Transport Satellite-1R} ({\it MTSAT-1R}) are used to derive yearly average clear fractions over the regions of Indonesia. This parameter is determined from temperature measurement of the IR3 channel (water vapor, 6.7 $\mu$m) for high altitude clouds (cirrus) and from the IR1 channel (10.7 $\mu$m) for lower altitude clouds. Accordingly, an algorithm is developed to detect the corresponding clouds. The results of this study are then adopted to select the best possible sites in Indonesia to be analysed
further by performing in situ measurements planned for the coming years. The results suggest that
regions of East Nusa Tenggara, located in south-eastern part of Indonesia, are the most promising
candidates for such an astronomical site. Yearly clear sky fraction of this regions may reach better than 70 per cent
with an uncertainty of 10 per cent. 
\end{abstract}


\section{Introduction}

Modern astronomical observatories are currently being planned and developed in many sites, as national observatory or
as joint international consortium involving some countries, across all continents. Large and sophisticated facilities, especially for a dedicated instrument and/or wavelength, must consider various excellent observing conditions. One of important factors that determine the success of developing and operating such astronomical facilities is the climate condition of the chosen site. The identification of potential sites are usually made after a careful long study. In Asia, we can see examples from Maidanak Observatory in Uzbekistan (Ehgamberdiev et al. 2000) and Devasthal in India (Sagar et al. 2000). 

Analysis of this climate factor could comprise several important characterisations, such as cloud cover, atmospheric thermal background, sky brightness, precipitable water vapour, atmospheric seeing and optical extinction, and so on. In North Africa, optical seeing at the Ouka\"{\i}meden has been long monitored (Siher \& Benkhaldoun 2004, Benkhaldoun et al. 2005). Not far from it in Spain, the largest European astronomical sites in the northern hemisphere, in La Palma, characterisations have been continuously monitored (Lombardi et al. 2006, 2007, 2008), and also in Calar Alto (S\'{a}nchez et al. 2007). In addition, Varela et al. (2008) analysed atmospheric aerosol intensively above Canary Island based on both satellite data and in situ measurements.
Site characterisation in San Pedro M\'{a}rtir, Mexico, has also been made after more than three
decades of its operation (Tapia et al. 2007). It also includes measurement of millimetric and infrared opacities.
In California, Barcroft Facility of the White Mountain Research Station is identified as an excellent site
for microwave observation (Marvil et al. 2006). 
Therefore, careful site selection has been extremely important to consider from the very beginning of the project.
Meanwhile, many major observatories continue to evaluate their sites to better understand the climate and character of the sites for observation scheduling and development planning. Site testing and selection for the Thirty Meter Telescope
(TMT) is another relevant example of modern astronomical site selection (Sch\"{o}ck et al. 2009 and references therein).

Presently, most major ground-based observatories are built either on coastal mountains, such as in Chile and 
in mid-west America, or on mountains in relatively small island, such as in Canary Island and in Hawaii. The altitude
of the mountains is higher than 2000 m. Most of them are located in subtropical zone, in latitude between 25$^\circ$
and 35$^\circ$ north or south. Efforts to search the best possible site on Earth has also led astronomers to survey the polar regions since they certainly could provide cold, dry, and dark winter skies with mild wind, excellent
seeing, and high clear sky fraction (see, for example, Andersen \& Rasmussen 2006; Saunders et al. 2009;
Steinbring et al. 2010). 

In this work, we present the search for a new astronomical site in Indonesia which is located around the equator. 
This country covers 95$^\circ$--141$^\circ$E and 6$^\circ$N to 11$^\circ$S. Hence, this represents a vast geographical area and entirely in tropical region. Note also
that Indonesia is an archipelago dominated by seas, dotted with a few big islands and a lot of small islands. This study is motivated by a plan to build a new
astronomical observatory, as an extension of the Bosscha Observatory, located in Lembang (107$^\circ$36$'$E, 6$^\circ$49$'$S), the only major observatory in Indonesia. Note that there were already plans in 1980s to propose a 2-m class telescope (van der Hucht 1984) and a consortium to build a giant equatorial radio telescope (Swarup, Hidayat \& Sukumar 1984) in Indonesia. However, they have yet to be realised. 

After more than eight decades, a new modern astronomical facility is thought necessary and is required to continue and advance astronomical research and education in Indonesia. The desired facility is a new modern one, in the framework of national observatory, operated within a concept of multi-wavelength observatory; primarily in the optical, infrared, and radio. This may represent a long-term program that must be continuously reviewed and pursued. Therefore, to realise such a facility, a series of preparatory work must be conducted. 

As outlined above, to determine the best possible locations for astronomical sites, there are many atmospheric and geographical parameters which must be studied carefully; and eventually site testing and monitoring must be performed for a long period of time, i.e., more than 3 to 5 years. Hence, first of all, we must identify certain regions that could provide a better meteorological condition. For this aim, we decide to evaluate the clear sky fraction, as the first important parameter, above Indonesia using meteorological satellite data from a period of more than a decade, to obtain the pattern and quantity of cloud coverage. We adopt methodology described in Erasmus \& Sarazin (2002). Similar study has been made for many astronomical observatories (see, for example, Erasmus \& van Rooyen 2006, Sebag et al. 2007, della Valle et al. 2010, Cavazzani et al. 2011). Motivation of this study and the use of satellite data are outlined in the following.

The paper is organized as follows. The satellite data bases used in this study is depicted in Section 2. In Section 3,
we describe the methodology adopted in this work. Furthermore, the data analysis is presented in Section 4, and is followed
by the site selection analysis in Section 5. Discussion of the results is given in Section 6, and finally conclusion is
drawn in Section 7.

\begin{table}
\centering
\caption{Data channels of meteorological satellite.}
\begin{tabular}{|c|c|} \hline
Channel & Wavelength ($\mu$m) \\ \hline
IR1 & 10.3 -- 11.3 \\
IR2 & 11.5 -- 12.5 \\
IR3 (water vapour) & 6.5 -- 7.0 \\
IR4 (NIR) & 3.5 -- 4.0 \\
VIS (albedo) & 0.55 -- 0.90 \\ \hline
\end{tabular}
\end{table}

\section[]{Satellite data}

As explained in Erasmus \& Sarazin (2002), meteorological satellites, particularly those in geostationary orbit are very useful sources of data for astronomical site characterization (see also Sarazin, Graham \& Kurlandczyk 2003). Moreover, the very high
stationary orbit ($\sim$36000 km) provides an extremely stable condition and is not influenced by the phenomena of
the high exosphere (Cavazzani et al. 2011).
Most meteorological satellites monitor cloud cover and other useful atmospheric parameters over large sections of the globe with a spatial and radiometric resolution suitable for evaluations at astronomical sites. Note that current satellite data archives cover 30 or more years of data of many areas, thus providing an adequate climatological data base for monitoring, comparison, and forecasting evaluation. 

With the availability of such data, cloud cover studies have been reported by many authors 
(Sagar et al. 2000, using {\it INSAT} satellite; Erasmus \& van Rooyen 2006, using {\it Meteosat} satellite; Sebag et al. 2007; della Valle et al. 2010; Cavazzani et al. 2011, using {\it GOES} satellite) and subsequently can be compared to the existing in situ measurement of the corresponding sites. As also
mentioned in Cavazzani et al. (2011), data analyses also provide homogeneous methodology since it is not dependent on
different judgement based on observer logbooks or on different instruments. Therefore, in our case,
since ground measurements of the possible sites, described in Section 4, are mostly not available yet, the use of satellite data
is indeed the most useful in guiding our initial site selection process.
This will be the primary source of our long term programs prior to in situ measurements.

As a part of our astroclimate studies to infer clear sky fraction above Indonesia, we use high spatial resolution data collected by several meteorological satellites. For this purpose, we consider data set from  http://weather.is.kochi-u.ac.jp/index-e.html, a weather data source site managed by the Kochi University, Japan\footnote{See also http://weather.is.kochi-u.ac.jp/wiki/archive/x\_e5\_bc\_ 95\_e7\_94\_a8\_e6\_96\_b9\_e6\_b3\_95}. The data set has been collected by the {\it GMS 5} for the period of 1995 to 2003, and {\it GOES 9} for the period of 2003 to 2005. Subsequently, starting from mid-2005, data were collected by {\it MTSAT-1R} satellite (also called {\it Himawari-6}), operated by Japan Meteorological Agency. The latter is recently replaced by {\it MTSAT-2} (also called {\it Himawari 7}) in July 2010. 

The area covered by these data set is from 70$^\circ$N to 20$^\circ$S in latitudes, and 70$^\circ$ -- 160$^\circ$E in longitudes in single image, thus covering the whole part of Indonesia. The images in these data set provide spatial resolution of 1/20$^\circ$, recorded in portable grey map (pgm) format, with an image size of 
1800 $\times$ 1800 pixels. It is equivalent to a spatial resolution of $\sim$5.6 km. A complete day coverage is provided with a temporal resolution of one hour. 

\begin{table}
\caption{Comparison of frequencies obtained from the sky visual inspection above Lembang vs three computations}
\begin{tabular}{lcccc}
\hline									
	&	Visual	&	Model 1*	&	Model 2*	&	Model 3*	\\
	&	obs.	&	(270)	&	(273)	&	(277)	\\
\hline									
Clear	&	278	&	312	&	291	&	255	\\
Mainly clear	&	51	&	46	&	40	&	35	\\
Mostly cloudy	&	176	&	151	&	157	&	163	\\
Cloudy	&	241	&	237	&	258	&	293	\\
\hline									
\hspace{0.5cm}$\chi^2$	&		&	8.45	&	7.03	&	19.65	\\
\hline									
\end{tabular}
Note: *Results of computation. Number in the bracket corresponds to temperature in K for $T_{10.7}$, i.e., the assumed threshold for mid-latitude clouds, while the assumed threshold for high cirrus adopted in this computation is $T_{6.7}=243$ K
(see text for more details).  
\end{table}

The satellite observations give five simultaneous images, recorded by instruments at five different channels of 
wavelength (one in visible and four in infrared), listed in Table 1. But note that the IR4 channel is made available by
{\it MTSAT} only. These channels may indicate the presence of clouds at different height. 
After evaluating the calibration (Tahara et al. 2004), there exists discrepancies in the observed brightness temperatures between {\it GMS 5} and {\it GOES 9} that need to be corrected for further analyses. However, there is a good agreement between {\it GOES 9} and {\it MTSAT-1R} and no particular systematic error is found (Tahara \& Ohkawara 2005). Moreover,
calibration table to convert graphics pixel values to brightness temperature is provided by Kochi University. 
Considering both the temporal resolution and the available wavelength channels,
consequently, the advantage of using these data is that we can make separate analysis for both day time and night
time conditions.

\section{Methodology}

Furthermore, we implement a simple algorithm for data analyses adopting the theory of cloud detection described in Erasmus \& Sarazin (2002) (see also, Rossow \& Garder 1993; Soden \& Bretherton 1996; Rossow \& Schiffer 1999). In this work, we only use two data channels: infrared channel IR1 (10.3--11.3 $\mu$m) and water vapor channel IR3 (6.5--7.0 $\mu$m), from the year 1996 to 2010. As can be shown from the corresponding weighting function (Erasmus \& Sarazin 2002; Cavazzani et al. 2011), the
latter channel may indicate the presence of high altitude cirrus clouds (higher than 8 km) and the former channel may
indicate the presence of mid-altitude clouds (at altitudes around 4 km). Atmospheric phenomena below this altitude,
such as fog and semi transparent cloud, could not be detected by this method. 

We further assume that radiation intercepted by the detector onboard the satellite has blackbody properties and is consequently described by Planck's law. Spectra of the received blackbody radiation are associated with temperatures of the top clouds that emitted the radiation. In our case, the data image with a size of 1800 $\times$ 1800 pixels have values between 0 and 255, accompanied by a data file to calibrate the pixel values into the brightness temperature in kelvin.

\begin{figure}
\epsfig{figure=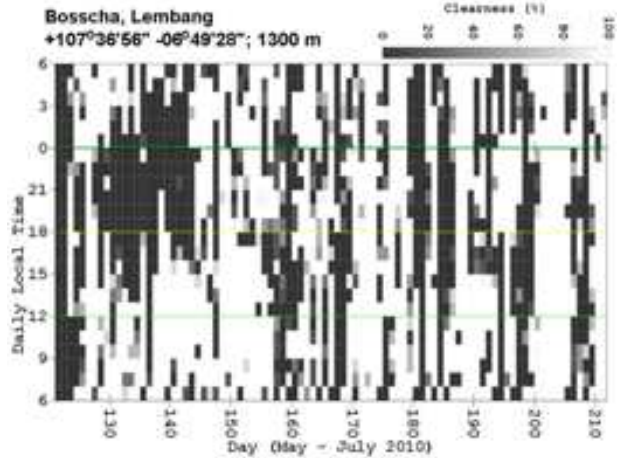,width=8.1cm}
 \caption{Daily chart of cloud detection, computed from Model 2 (see Table 2) that reproduces better the sky visual inspection above Lembang at a temporal resolution of 1 hour, during May to July 2010. The blacker the chart the more opaque the sky is (see text for more details).}
\end{figure}

In clear sky condition, the emitted electromagnetic radiation at the IR1 channel is practically not absorbed by the atmosphere, so the radiation measured by the satellite originates from the surface (land or ocean). However, in the presence of cloud, the radiation is absorbed by the cloud and is reemitted as a blackbody radiation at infrared wavelength. Consequently, the presence of cloud gives the effect as if from a `higher surface', since the detected radiation originates from the cloud top. On the other hand, radiation from the IR3 channel intercepted by the detector is affected by absorption and re-emission by water vapours. These water vapours exist in an atmospheric layer between 600 mbar and 300 mbar pressure level.  
Therefore, radiation from this layer strongly depends on the content of water vapor and its temperature. 

Erasmus \& Sarazin (2002) used an empirical relation between upper troposheric humidity ($UTH$) with brightness temperature that is measured in the IR3 channel ($T_{6.7}$):
\begin{equation}
UTH = \frac{\cos \theta}{p_0} e^{31.5-0.115 T_{6.7}}
\end{equation}
where $\theta$ is the satellite's zenith angle, and $p_0 = p[T=240 K]/300$ mbar is a reference pressure level 
($p_0 \sim 0.9$ for tropical regions, Soden \& Bretherthon 1996). With this relationship, $UTH$ can be determined from the temperature measured by the satellites. 

\begin{figure*}
  \centering
  \epsfig{figure=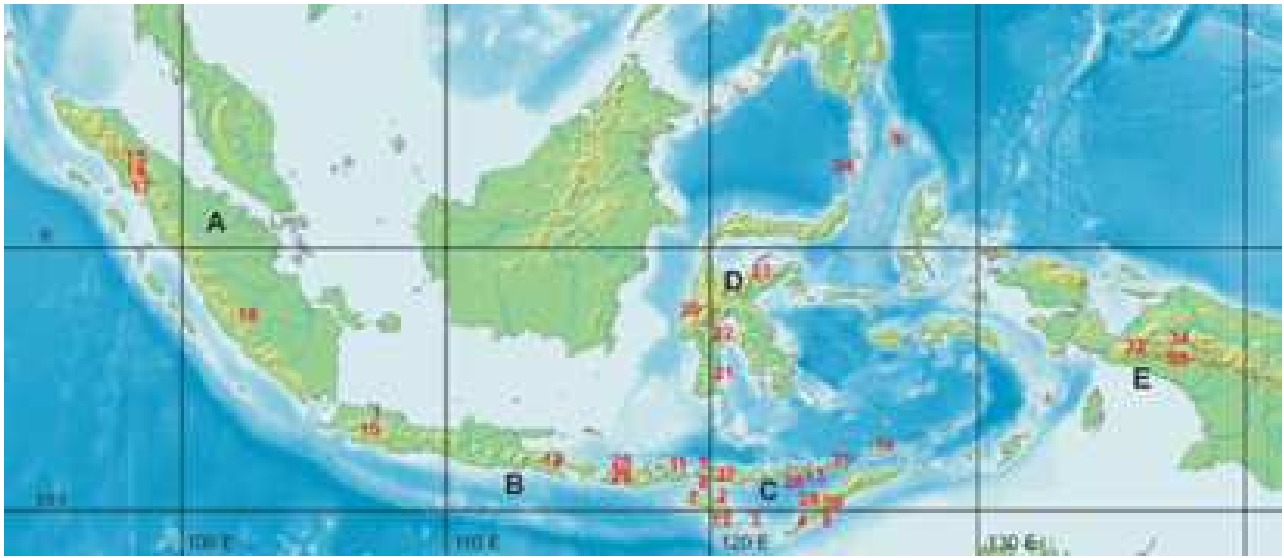,width=17.cm}
  \caption{Map of Indonesia: (A) Sumatra, (B) Java, and isles to the east are Bali and West Nusa Tenggara,
(C) East Nusa Tenggara, (D) Sulawesi, and (E) Papua. Identification numbers displayed on the map indicate approximate positions
of the sites mentioned in Table \ref{site}, to be analysed in this work (see text for details).}
\end{figure*}

To detect the presence of clouds in a pixel, first we consider temperature $T_{6.7}$ from the IR3 image. The presence of cirrus clouds provides very low $T_{6.7}$, or equivalently, very high $UTH$ ($>100$ per cent). Hence, a threshold value 
for $T_{6.7}$ must be determined. If the IR3 channels show the presence of clouds, then we have $T_{6.7} < T_{6.7,threshold}$ or $UTH > UTH_{threshold}$, and we identify the corresponding pixel as cloudy. All the threshold value is actually determined by
comparison with zeroth order ground data, described in the following section.

Subsequently, if the IR3 channels do not detect the presence of clouds, we still have to inspect the data from the IR1 channel that provide temperature we denote $T_{10.7}$. 
As described previously, the corresponding radiation is associated to the altitudes of the emitter that can be land, ocean, or cloud. To detect this mid-altitude cloud, a threshold value must also be determined. 
If the temperature measured by the detector is greater than this value, we consider that the radiation originates from surface, land or ocean, that is, no cloud is detected. Otherwise, the mid-altitude cloud is present. Hence, the presence of cloud is determined from two successive steps. 

Moreover, for the site area analysis, note that
a pixel represents an area of $\sim$5 km $\times$ 5 km on the Earth's surface. Therefore, in this work, to inspect the cloud cover of a region, we take samples of an area of 5 pixel $\times$ 5 pixel, centered at a location of interest corresponding approximately to an area of observations within $\sim$$50^\circ$ zenith angles. 
To identify the observing night condition from this site area analysis, we adopt the same definition used in Erasmus \& Sarazin
(2002). {\it Clear} or {\it photometric} nights correspond to all 25 pixels clear above the analysed site. 
Furthermore, {\it transitional} or {\it spectroscopic} nights correspond to at least 20 of 25 pixels clear. When the number of clear pixels are less than 20, the observing condition is considered {\it opaque}. Subsequently,
for each hour data, two images (IR1 and IR3) for the whole Indonesia are scanned using the method described above and we derive the percentage of the clear sky fraction over the region under study. 

\section{Data analysis}

\subsection{Data validation and error estimation}

Validation of the satellite data is not an easy task. For example, Khaiyer et al. (2004) show validation of the {\it GOES 9} derived products using ground-based instrumentation over the tropical western Pacific region. In our analysis, we compare the computational results, retrieved from the satellite data, corresponding to the site of the Bosscha Observatory in Lembang with the observer log book. However, this comparison is very limited to night time only, and lack of completeness due to down time, mainly because of bad weather condition and rainfall situation, which is not regularly noted in the log book. Therefore, we decide to conduct the sky visual inspection of clouds above Lembang including day time to obtain the daily coverage of the clouds. 

\begin{figure*}
  \centering
\vbox{
\centerline{\epsfig{figure=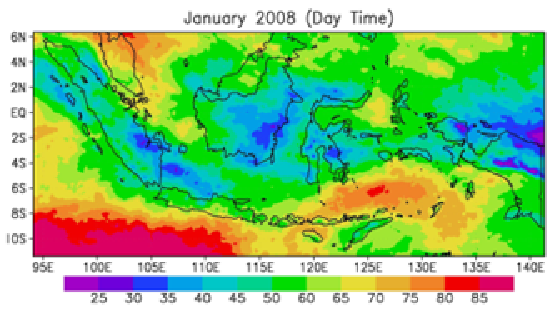,height=36mm}\hspace{3mm}\epsfig{figure=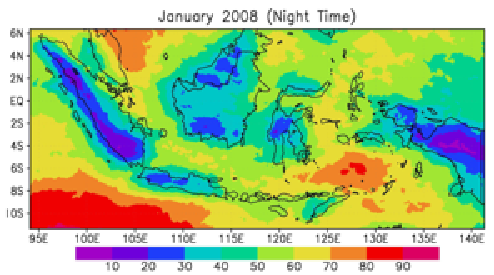,height=36mm}}
\vspace{1mm}
\centerline{\epsfig{figure=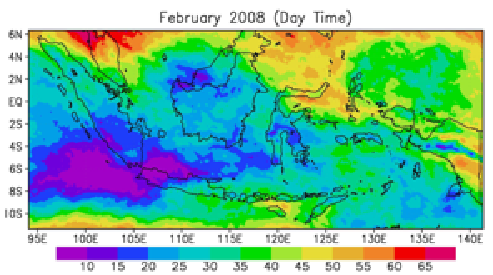,height=36mm}\hspace{3mm}\epsfig{figure=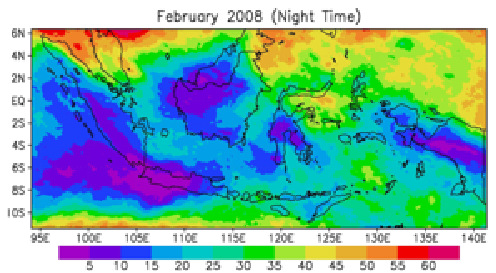,height=36mm}}
\vspace{1mm}
\centerline{\epsfig{figure=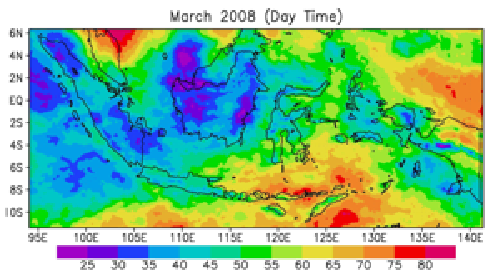,height=36mm}\hspace{3mm}\epsfig{figure=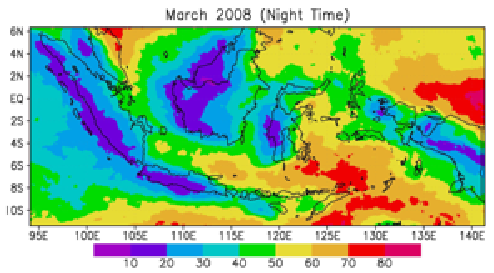,height=36mm}}
\vspace{1mm}
\centerline{\epsfig{figure=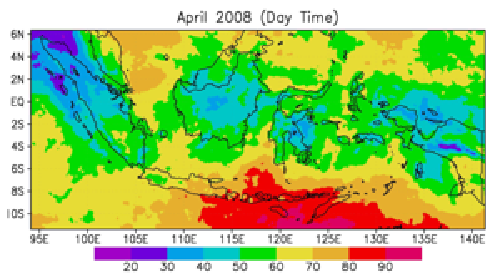,height=36mm}\hspace{3mm}\epsfig{figure=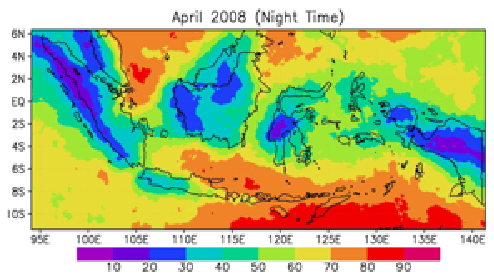,height=36mm}}
\vspace{1mm}
\centerline{\epsfig{figure=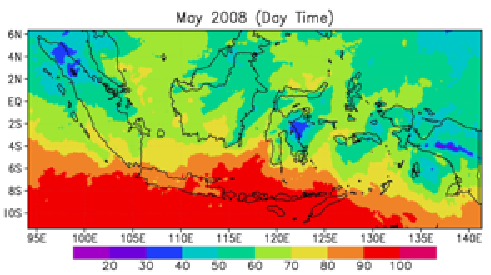,height=36mm}\hspace{3mm}\epsfig{figure=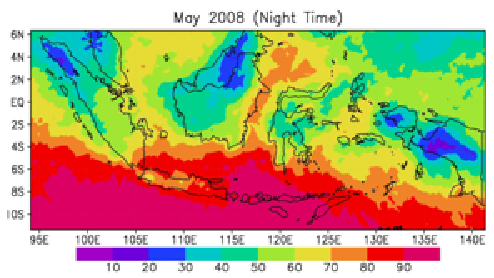,height=36mm}}
\vspace{1mm}
\centerline{\epsfig{figure=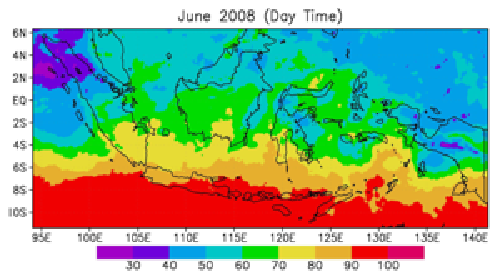,height=36mm}\hspace{3mm}\epsfig{figure=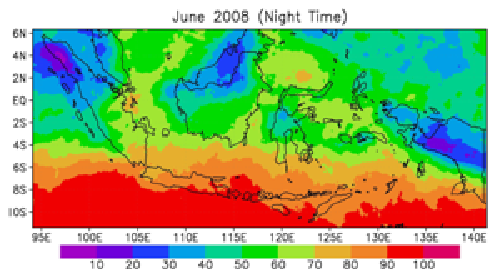,height=36mm}}
}
\caption{Monthly average clear sky fraction in Indonesia from January to June 2008 for day time (left) and night
time (right), expressed in percentage, for comparisons. Note that the scale in the legend is not the same for each figure.
During the first three months, most land regions in Indonesia are very cloudy. Starting in April, the southern part is clear.}
\end{figure*}
\begin{figure*}
  \centering
\vbox{
\centerline{\epsfig{figure=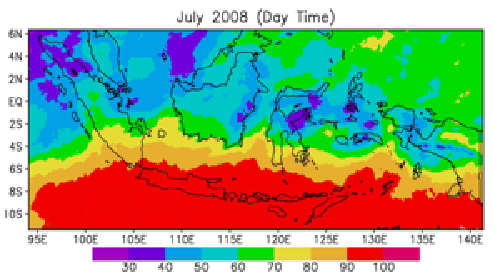,height=36mm}\hspace{3mm}\epsfig{figure=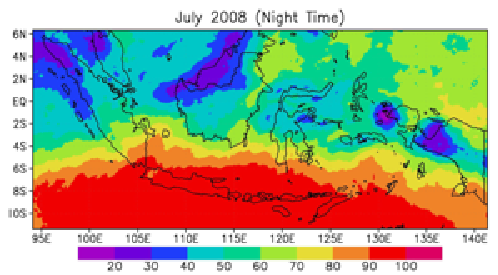,height=36mm}}
\vspace{1mm}
\centerline{\epsfig{figure=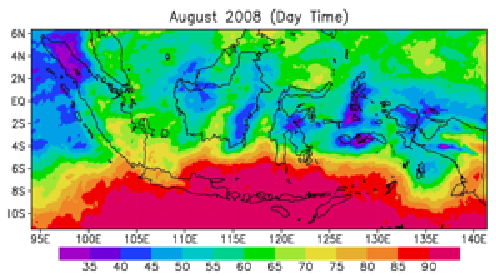,height=36mm}\hspace{3mm}\epsfig{figure=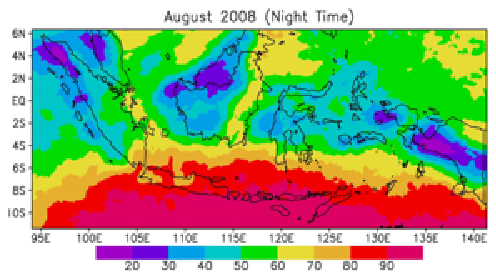,height=36mm}}
\vspace{1mm}
\centerline{\epsfig{figure=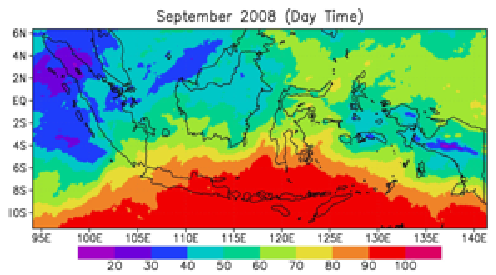,height=36mm}\hspace{3mm}\epsfig{figure=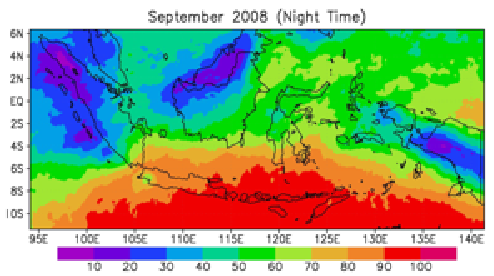,height=36mm}}
\vspace{1mm}
\centerline{\epsfig{figure=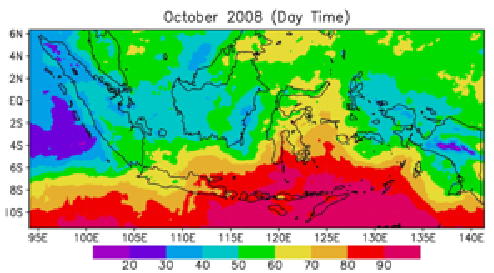,height=36mm}\hspace{3mm}\epsfig{figure=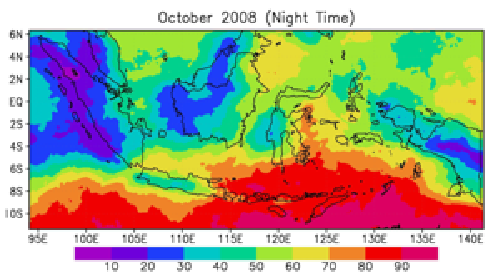,height=36mm}}
\vspace{1mm}
\centerline{\epsfig{figure=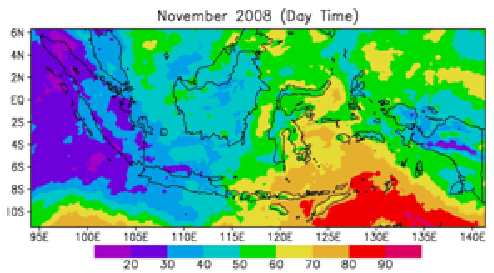,height=36mm}\hspace{3mm}\epsfig{figure=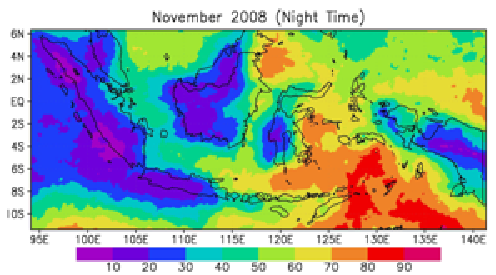,height=36mm}}
\vspace{1mm}
\centerline{\epsfig{figure=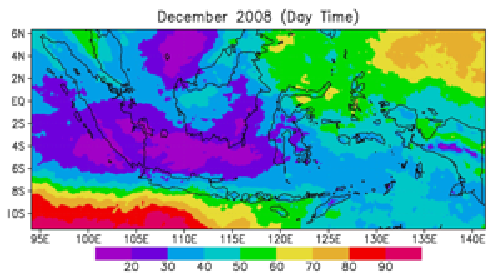,height=36mm}\hspace{3mm}\epsfig{figure=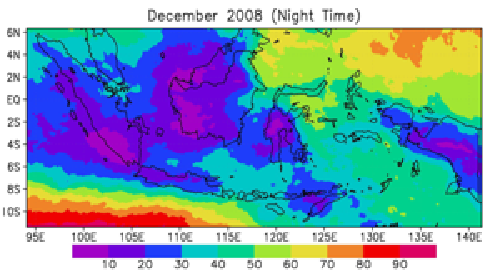,height=36mm}}
}
\caption{Monthly average clear sky fraction in Indonesia from July to December 2008 for day time (left) and night
time (right), expressed in percentage. Note that the scale in the legend is not the same for each figure.
The opposite situation of Fig.~3 is found. Very cloudy situation peaked in December 2008. }
\end{figure*}

\begin{figure*}
\centering
\vbox{
\centerline{\epsfig{figure=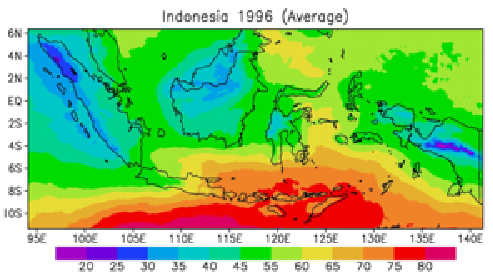,width=59mm}\hspace{1mm}\epsfig{figure=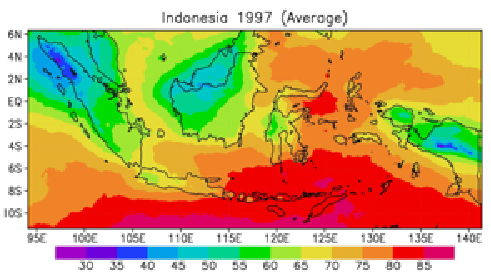,width=59mm}\hspace{1mm}
\epsfig{figure=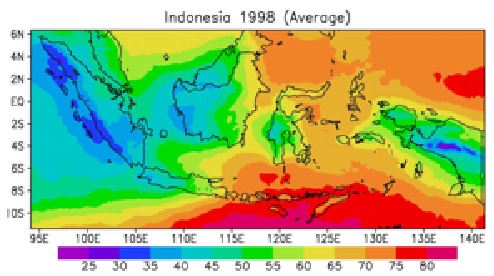,width=59mm}}
\vspace{1mm}
\centerline{\epsfig{figure=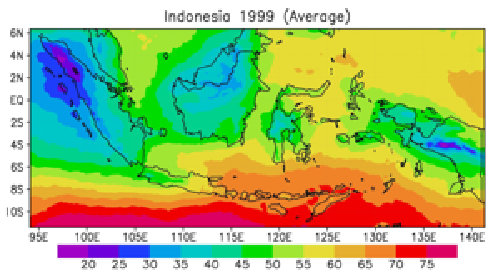,width=59mm}\hspace{1mm}\epsfig{figure=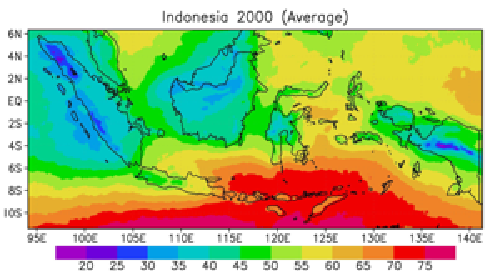,width=59mm}\hspace{1mm}\epsfig{figure=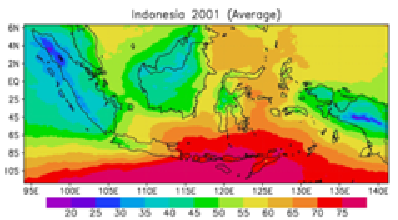,width=59mm}}
\vspace{1mm}
\centerline{\epsfig{figure=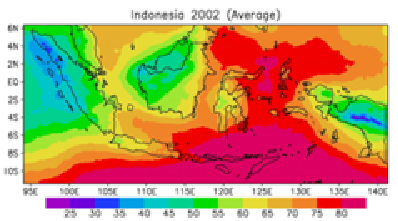,width=59mm}\hspace{1mm}\epsfig{figure=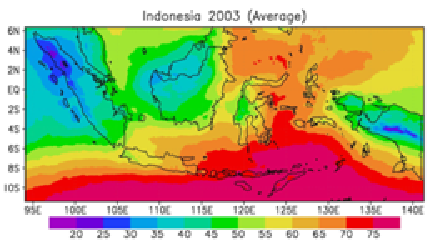,width=59mm}\hspace{1mm}\epsfig{figure=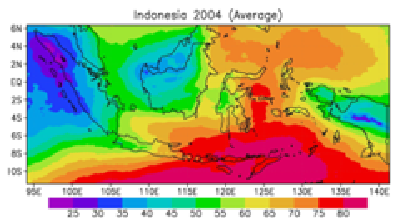,width=59mm}}
\vspace{1mm}
\centerline{\epsfig{figure=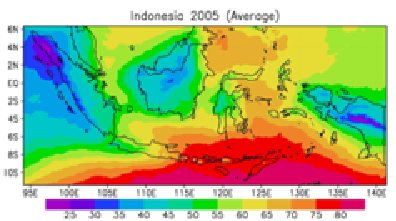,width=59mm}\hspace{1mm}\epsfig{figure=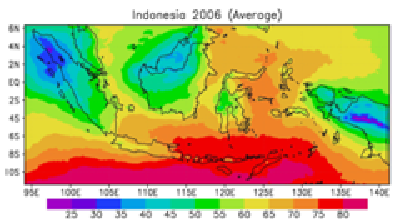,width=59mm}\hspace{1mm}\epsfig{figure=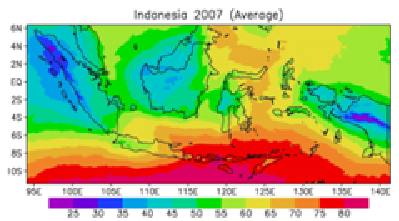,width=59mm}}
\vspace{1mm}
\centerline{\epsfig{figure=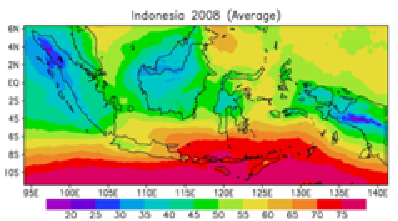,width=59mm}\hspace{1mm}\epsfig{figure=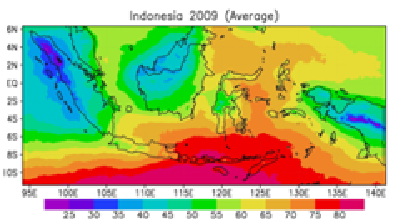,width=59mm}\hspace{1mm}\epsfig{figure=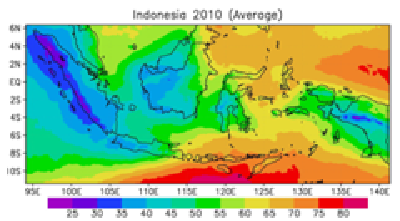,width=59mm}}
}
\caption{Yearly average clear sky fraction in Indonesia from 1996 to 2010. Note that the scale in the legend is not the same for each figure. We see that south-eastern part is consistently clear.}
\end{figure*}

\begin{figure*}
\centering
\vbox{
\centerline{\epsfig{figure=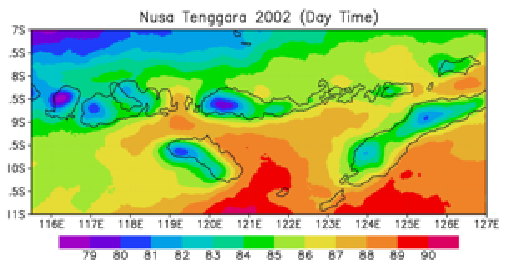,width=8.0cm}\hspace{2mm}\epsfig{figure=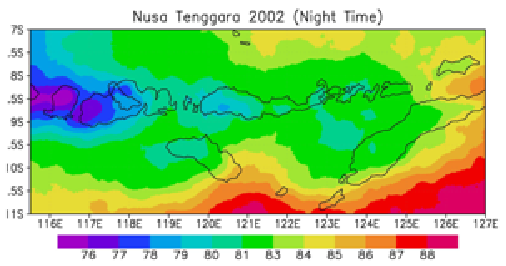,width=8.0cm}}
\vspace{2mm}
\centerline{\epsfig{figure=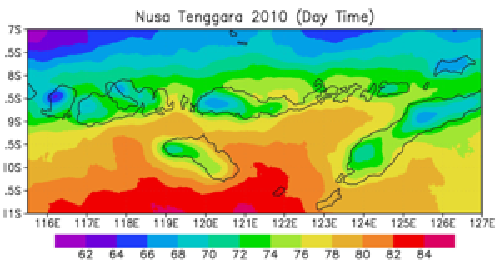,width=8.0cm}\hspace{2mm}\epsfig{figure=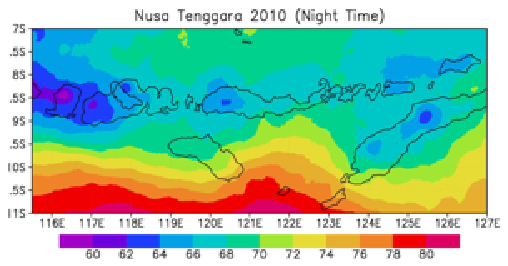,width=8.0cm}}
\caption{Yearly average clear fraction in Nusa Tenggara for 2002 (the best) and 2010 (the worst) both in day time and night time for comparison. Note that the scale in the legend is not the same for each figure. Notice again that the eastern part
is clearer than the western part at about 5 per cent.}
}
\end{figure*}

We observed systematically the appearance of clouds above the Bosscha Observatory qualitatively every hour corresponding to the temporal resolution of the satellite data. Almost every day, between May and July 2010, we sampled hourly the sky, although
most of them were not a 24 hours coverage. Following Steinbring et al. (2010), we classify our visual inspection in four bins: clear, mainly clear (when more than half of the sky is clear), mostly cloudy (when more than half of the sky is cloudy), and cloudy. We acquired 746 frequencies shown in Table 2. Note however that this qualitative visual inspection depends on observer's judgement. Cirrus are usually not visible from the ground especially during the night. They are also not generally distributed uniformly over the sky. 
The agreement between models and the sky visual inspection in Table 2 can be estimated to be roughly 90 per cent. However if we consider only the night time samples (256 frequencies), the agreement is worse, roughly 50 per cent. 
Therefore, absolute calibration is beyond the scope of this study and we estimate that the accuracy of this visual inspection cannot be better than 10 per cent. 

Subsequently, we consider the tropical atmosphere model (Kneizys et al. 1996) and adopt the temperature values at the corresponding altitudes. We note that the computational result may be sensitive to the threshold values. A slightly different value can yield different results. To determine the corresponding threshold temperatures, we assumed three
temperature values: 238, 243, 248 K for high cirrus clouds (altitudes between $\sim$8000--10000 m)
and 270, 273, and 277 K for mid-altitude clouds (Model 1 to 3, respectively; altitudes between $\sim$4000--5000 m). We found for high cirrus that the cloud fraction is not sensitive in this assumed temperature range, but it is not the case for mid-altitude clouds. Table 2 shows the results of the computation as compared to the sky visual inspection. A simple $\chi^2$ analysis leads that $T_{6.7} = 243$ K and $T_{10.7} = 273$ K for IR3 and IR1 channels, respectively, present the best possible values, giving the smallest $\chi^2$ value of 7.03, among the three models as shown in Table 2. 
Fig.~1 shows a `daily chart', i.e., the appearance of hourly cloud in a complete day (24 hours) coverage as a daily time series, obtained from Model 2 (see Table 2). 

Note that there is always discrepancy between the satellite data and our visual inspection. The sources of this discrepancy are: (1) most local cloud observed on the site is too small to be resolved by the satellite imager.  On the other hand, (2) the area analysed is about 25 km$^2$ centered on the site, which is not easily quantified by the local observer. Low or mid-altitude clouds can affect either partial areas of the sky or its totality. (3) There are also 
missing data in the satellite observations that we recognize to occur randomly. In our computation, we consider all the
missing data as opaque. After carefully comparing the satellite data with our visual inspection, and considering the above most
possible sources of errors, we estimate that all these errors are about 10 per cent. 

We use the inferred values of threshold temperatures to compute the clear sky fraction for any location
since the assumed homogeneity of satellite data is justified. Moreover, we apply the same values to the whole data in the 
time span analysed in this work. It means that we apply the same parameters to the four satellites functioning in
different periods. This assumption may be inherently inaccurate and thus can introduce biases when different seasons and sites located at different altitudes are compared. However, for the purpose of this first study, the results are considered sufficient, accounting for the uncertainty that cannot be better than 10 per cent.
With all these limitations, we proceed to perform relative site comparisons described in Section 5.

\subsection{Monthly average clear fraction}

Fig.~2 shows the map of Indonesia, an archipelago straddling the equator, with more land mass situated in
south of the equator. Five regions denoted A (Sumatra), B (Java, Bali, and West Nusa Tenggara), C (East Nusa
Tenggara), D (Sulawesi), and E (Papua) are indicated on the map. Identification numbers 
corresponding to locations detailed in Table 3 are for later use in Section 5. 

The implemented algorithm is then used to scan the whole data for Indonesia to read the corresponding brightness
temperature at a time span of 15 years, from 1996 to 2010. The temporal resolution as well as the wavelength channels of the data permit us to analyse the clear sky fraction separately for day time (for the Sun observations and radio application) and night time (for all wavelength of interest from the radio to optical). 
For the first evaluation of the clear sky over Indonesia, we show in Fig.~3 and 4 the results of monthly average clear
sky fraction for day time and night time for comparison. We take the year 2008 that we consider as `normal' year
and we notice the seasonal cycle for a year. Obviously, the corresponding cycle may be heavily related to the phenomenon
of Asian monsoon that affects considerably across Indonesia. However, it is beyond the scope of the present work to analyse such relations.

Fig.~3 illustrates the difference in detail between monthly day time and night time clear fraction from January to June 2008.
The legend below the picture does not indicate the same scale. However, one may easily notice that the bluer the pattern,
the more cloudy the region is; and conversely, the redder the pattern, the clearer the region is. The difference between day and night time fluctuates and sometimes can be so pronounced as we can see for the month of February. The first three months in the year is usually marked by wet season in the south and dry season in the north. However, we note that from the equator up to latitude of 4$^\circ$N, the clearer sky does not manifest itself significantly. The situation changes in April where we see a much clearer sky in the south eastern part of Indonesia, with clear fraction about 70 to 80 per cent.

The situation for the following five months is displayed in Fig.~3 and 4. Clear fraction greater than 80 per cent appears 
persistently for regions below the latitude of roughly 5$^\circ$S. The difference between day time and night time
is not significant as well. Java is in general very clear during this period. The sky over the south east islands is 
even clearer. In July, the clear fraction is greater than 90 per cent. On the contrary, North Sumatra and Papua do 
not show a clear sky condition. Kalimantan (Borneo) has about 40 to 50 per cent, a slightly larger clear sky fraction compared to Sumatra. 

Transition to less clear condition occurs during October as shown in Fig.~4. Bad weather condition is persistent 
in November and mostly in December. The sky is almost completely cloudy and the clear sky fraction is lower than 20 per cent.
Similar analyses are made for all the data and we found that this general pattern is usually applicable, except for temporal
anomaly. 

\subsection{Yearly average clear fraction}

Subsequently, we present yearly average clear fraction above Indonesia for the past 15 years. 
We do not distinguish between day time and night time clear fraction to see the regional pattern 
as indicated by monthly mean clear fraction. 
Fig.~5 shows scan results of the fraction of clear sky for the 15 years, from 1996 to 2010 for comparison to each other. 
We clearly identify that south-east region of Indonesia, primarily East Nusa Tenggara (Region C in Fig.~2) has consistently
clear fraction larger than any other region, with values mostly greater than 70 per cent. Again note that these clearer regions are mostly located below the latitude of 5$^\circ$S where the land part constitutes of small islands. A bigger island like Java is also located at this latitude, but only eastern part has large clear fraction, between 65 to 70 per cent. West Java has clear fraction lower than 55 per cent. Bigger islands such as Sumatra, Kalimantan, and Papua have even lower value. Low values, less than 30 per cent, are found mainly in North Sumatra and mid-Papua that unfortunately coincide with high mountain areas (see Section 5).

Within the eight years, we find slight different patterns, mainly in 1997 and 2002, during which high clear 
fraction ($\sim$70 per cent) extends to equatorial regions between Sulawesi and Papua, in the archipelago of Maluku (Molucca Island). This also affects regions of North and South Sulawesi. A milder situation in the same regions was found in the following year (1998 and 2003, respectively) with a clear fraction of $\sim$65 per cent.

Comparison of 2002 with the following seven years from 2003 shows that they are not clearer. The regions with high value of clear fraction seems decreasing from year to year. The clearest region is always in the south east with the clear fraction greater than 70 per cent. Note that the year 2010 marked the regions with the less clear fraction for the past 15 years. We denote that year having extreme weather as we also observed at the Bosscha Observatory. The clear regions of Nusa Tenggara also show decreasing clear area with values between 60 to 65 per cent. In contrast, in the far west (Sumatra) and far east (Papua), we notice high mountains regions always have less clear fraction (less than 35 per cent).   
The highest summit in Papua, for example, has a large cloud coverage. Hence, very high mountains are not always advantageous for astronomy in this tropical region. Nevertheless, this will be confirmed by further analysis in Section 5.

Referring to these results, actually site selection can thus be focused in the south-east regions of Indonesia.
Fig.~6 shows the yearly average clear fraction for this region, called Nusa Tenggara, in 2002 and 2010 for
comparison both in day time and night time. We estimate qualitatively that the year 2002 had the largest area with
high clear sky fraction, while the year 2010 had the opposite situation. The presence of dense cloud contour on some islands with the shaded dark blue (Fig.~6) can be immediately associated with high mountains on these islands.
In 2002, West Nusa Tenggara has slightly less clear fraction ($\sim$75--80 per cent) compared to the East Nusa Tenggara ($\sim$80--85 per cent). In 2010, the situation is similar but the clear fraction is lower ($\sim$60--70 per cent).

As mentioned above, a pixel represents an area of $\sim$5 km $\times$ 5 km on the ground. Therefore, we can proceed further by
selecting a position of interest and inspect an area of 5 pixel $\times$ 5 pixel to see how the clear fraction evolves hourly
for a given year, centered in the selected location. For this, we select some reference sites to be
compared. It is described in the following section. 

\section{Comparative site analysis}

\subsection{Site selection criteria}

Guided by the results of clear sky fraction for the whole Indonesia presented in the previous section, we could select
a number of locations around the country which are expected to be suitable to host an astronomical observatory.
The general criteria for this selection are: (1) located at sufficiently high altitude, preferably higher than 2500 m;
(2) far enough from dense population, preferably in national park with special permission; (3) reasonably accessible; 
(4) proximity to the possible availability of infrastructure, such as electricity and roads. Otherwise, power plant is
preferably provided using renewable system.

For completeness, the selection is not limited to the region of Nusa Tenggara only, but still also considers a number of other locations that satisfy our general criteria. This leads us to divide our selection to
five different regions (A to E, Fig.~2). This selection also envisages regions in the north, around, and south 
of equator. Despite our preference of south regions, in any case if available, sites close to the equator could be also
interesting since they offer the same coverage of both northern and southern hemisphere.
Sumatra (A) is analysed because it has many high mountains, and a number of them are situated in the north of equator.
In contrast, Kalimantan, a big island in the east of Sumatra, is not analysed since it has no high altitudes, and as shown earlier, the sky clear fraction in this region is generally not promising.

\begin{table*}
\begin{center}
\caption{The 34 sites in Indonesia analysed in this work and the corresponding clear sky fraction for 2006 to 2010}\label{site}
\begin{tabular}{clcccccccccc}																						
\hline\hline																						
No	&	\mbox{}\hspace{0.7cm} Site	&	Region$^\dagger$ & Longitude 	&	Latitude	&	Alt. 	&	\multicolumn{6}{c}{				Clear sky fraction (per cent)}							\\	\cline{7-12}
	&		&	& (East)	&		&	(m)	&	2006	&	2007	&	2008	&	2009	&	2010	&	Mean*	\\	\hline
	&	{\it Sea-level}** 	& &		&		&		&		&		&		&		&		&		\\	
1	&	Jatiwangi	&	B & $108^\circ 10' 1''$	&	$-6^\circ  46' 4''$	&	45	&	59.9	&	51.9	&	48.9	&	51.7	&	31.8	&	48.8	\\	
2	&	Waingapu	&	C & $120^\circ 16' 1''$	&	$-9^\circ 39' 46''$	&	20	&	72.9	&	75.0	&	71.0	&	70.5	&	67.0	&	71.3	\\	
3	&	Kupang***	&	C & $123^\circ 34' 40''$	&	$-10^\circ 10' 19''$	&	25	&	73.2	&	73.8	&	69.8	&	70.4	&	66.5	&	70.7	\\	
	&		&		&		&		& &		&		&		&		&		&		\\	
	&	{\it Low-altitude}** 	&	&	&		&		&		&		&		&		&		&		\\	
4	&	Sawu***	&	C & $121^\circ 52' 38''$	&	$-10^\circ 33' 6''$	&	320	&	75.4	&	78.4	&	74.1	&	73.6	&	74.2	&	75.1	\\	
5	&	Komodo	&	C & $119^\circ 26' 51''$	&	$-8^\circ 33' 52''$	&	800	&	72.5	&	74.5	&	70.0	&	69.8	&	63.9	&	70.1	\\	
6	&	Rinca	&	C & $119^\circ 40' 19''$	&	$-8^\circ 45' 16''$	&	700	&	71.7	&	72.3	&	69.1	&	68.3	&	63.4	&	69.0	\\	
7	&	South Tambolaka & C	&	$119^\circ 16' 28''$	&	$-9^\circ 35' 26''$	&	878	&	70.8	&	72.3	&	68.4	&	66.7	&	61.7	&	68.0	\\	
8	&	East Kupang	&	C & $123^\circ 59' 18''$	&	$-10^\circ 15' 43''$	&	665	&	72.8	&	73.7	&	70.6	&	70.6	&	66.4	&	70.8	\\	
9	&	Talaud	&	D & $126^\circ 49' 33''$	&	$+4^\circ 18' 38''$	&	640	&	56.4	&	55.4	&	45.3	&	52.2	&	58.4	&	53.5	\\	
	&		&		&		&		& &		&		&		&		&		&		\\	
	&	{\it Mountain sites}	& &		&		&		&		&		&		&		&		&		\\	
10	&	Lembang***	&	B & $107^\circ 36' 56''$	&	$-6^\circ 49' 28''$	&	1310	&	56.4	&	47.8	&	46.9	&	48.6	&	29.1	&	45.8	\\	
11	&	East Sumbawa	&	B & $118^\circ 55' 27''$	&	$-8^\circ 28' 56''$	&	1419	&	72.2	&	75.0	&	70.6	&	69.3	&	62.4	&	69.9	\\	
12	&	South Waingapu 	&	C & $120^\circ 14' 11''$	&	$-10^\circ 6' 56''$	&	1200	&	71.3	&	74.3	&	69.3	&	70.5	&	67.5	&	70.6	\\	
13	&	Sirung	&	C & $124^\circ 6' 19''$	&	$-8^\circ 31' 7''$	&	1310	&	72.7	&	71.6	&	67.6	&	68.0	&	63.5	&	68.7	\\	
14	&	Wetar	&	C & $126^\circ 30' 47''$	&	$-7^\circ 41' 9''$	&	1390	&	70.3	&	69.3	&	65.2	&	65.1	&	57.5	&	65.5	\\	
15	&	Sinabung	&	A & $98^\circ 23' 77''$	&	$+3^\circ 10' 19''$	&	2400	&	23.9	&	24.2	&	19.0	&	19.7	&	21.3	&	21.6	\\	
16	&	Sibayak	&	A & $98^\circ 25' 58''$	&	$+2^\circ 55' 10''$	&	2400	&	22.6	&	23.4	&	19.1	&	20.2	&	20.7	&	21.2	\\	
17	&	Silimapuluh	&	A & $98^\circ 44' 56''$	&	$+2^\circ 27' 23''$	&	2100	&	24.1	&	24.9	&	20.3	&	21.9	&	21.9	&	22.6	\\	
18	&	Kerinci	&	A & $101^\circ 51' 28''$	&	$-2^\circ 30' 8''$	&	2900	&	39.6	&	36.2	&	34.9	&	36.0	&	35.9	&	36.5	\\	
19	&	West Ijen	&	B & $114^\circ 8' 10''$	&	$-8^\circ 6' 45''$	&	2340	&	66.1	&	63.2	&	59.5	&	61.9	&	48.0	&	59.7	\\	
20	&	Tibo	&	D & $119^\circ 22' 22''$	&	$-2^\circ 44' 53''$	&	3000	&	45.8	&	38.4	&	32.8	&	41.0	&	33.5	&	38.3	\\	
21	&	Lombosang	&	D & $119^\circ 56' 48''$	&	$-5^\circ 18' 48''$	&	2700	&	57.5	&	49.4	&	45.8	&	53.1	&	38.1	&	48.8	\\	
22	&	Rantemario	&	D & $120^\circ 2' 26''$	&	$-3^\circ 24' 33''$	&	3400	&	45.8	&	37.4	&	31.6	&	41.4	&	29.2	&	37.1	\\	
23	&	Binohoe	&	D & $122^\circ 11' 27''$	&	$-1^\circ 3' 50''$	&	2518	&	54.4	&	46.6	&	38.9	&	48.9	&	46.3	&	47.0	\\	
24	&	Sangihe-Siau	&	D & $125^\circ 24' 27''$	&	$+2^\circ 46' 54''$	&	1770	&	62.4	&	59.1	&	48.4	&	56.2	&	58.9	&	57.0	\\	
25	&	Rinjani 1	&	B & $116^\circ 27' 36.3''$	&	$-8^\circ 24' 44.9''$	&	3520	&	69.3	&	70.0	&	66.2	&	67.1	&	58.5	&	66.2	\\	
26	&	Rinjani 2	&	B & $116^\circ 34' 7.9''$	&	$-8^\circ 25' 40.6''$	&	2157	&	70.5	&	71.2	&	67.5	&	68.0	&	59.4	&	67.3	\\	
27	&	Ruteng	&	C & $120^\circ 31' 18''$	&	$-8^\circ 38' 12''$	&	2100	&	67.5	&	67.7	&	65.2	&	63.9	&	57.5	&	64.4	\\	
28	&	Lembata	&	C & $123^\circ 23' 3''$	&	$-8^\circ 32' 38''$	&	1590	&	71.5	&	71.9	&	68.1	&	67.7	&	63.7	&	68.6	\\	
29	&	Timau	&	C & $123^\circ 56' 15''$	&	$-9^\circ 34' 46''$	&	1685	&	70.6	&	70.1	&	66.9	&	67.2	&	60.9	&	67.1	\\	
30	&	Mutis	&	C & $124^\circ 13' 43''$	&	$-9^\circ 33' 39''$	&	2347	&	69.5	&	69.1	&	66.1	&	66.9	&	60.8	&	66.5	\\	
31	&	Alor	&	C & $125^\circ 1' 12''$	&	$-8^\circ 13' 14''$	&	1760	&	72.5	&	70.6	&	67.6	&	66.9	&	61.0	&	67.7	\\	
32	&	Erekebo	&	E & $135^\circ 51' 36''$	&	$-3^\circ 51' 15''$	&	3700	&	23.9	&	21.9	&	17.6	&	22.9	&	24.6	&	22.2	\\	
33	&	Puncak Jaya 1	&	E & $137^\circ 9' 23''$	&	$-4^\circ 4' 37''$	&	4700	&	10.6	&	9.3	&	7.4	&	10.8	&	13.7	&	10.4	\\	
34	&	Puncak Jaya 2	&	E & $137^\circ 18' 41''$	&	$-3^\circ 45' 27''$	&	4020	&	22.6	&	22.3	&	20.1	&	22.2	&	27.7	&	23.0	\\	\hline
\end{tabular}	\\
\underline{Notes:} $^\dagger$indicated in Fig.~2; *mean of 5 years; **for comparisons; ***sites for reference										
\end{center}
\end{table*}

There are not many choices for Java (B) according to criterion 2. This island is the most populated region in
Indonesia. High mountains (as active volcanoes) are generally close to settlement with dense population. However,
criteria 3 and 4 are usually fulfilled. Next to it to east, West Nusa Tenggara could have promising sites. 
Moreover, East Nusa Tenggara (C) must be naturally explored intensively, referring to the results of monthly and yearly
average clear sky fraction analysed previously. Note also that Sulawesi (D) has many high mountains. As we have seen from
Fig.~5, North and South Sulawesi may have large clear sky fraction. Finally, Papua (E) must also be surveyed since the
highest summit in Indonesia is located in this island.

Using the Google Earth, we have selected 34 locations, divided in three groups, i.e., (i) {\it sea level}: 3 locations, (ii) {\it low-altitude}: 6 locations, (iii) {\it mountain sites}: 25 locations (see Table 3). The corresponding number of the site is approximately shown in the map in Fig.~2. We also define a reference site for each group based on which we decide any selected site is either better or worse. For sea-level group, we take Kupang (Site 3) as reference. We have visited
this location several times and made preliminary measurement of seeing that we found to be better than in Lembang (Site 10),
in West Java, the site of  the Bosscha Observatory. Moreover, we represent Lembang as reference for the group of mountain site since we know its climate better and we are determined that the future site must be better than this site. Jatiwangi (Site 1) in West Java is analysed because of its properties as relatively dry place of the region. 

Sawu (Site 4), a small island, is also taken as reference for low-altitude site due to its unique position in the middle of
Sawu Sea, between Timor and Sumba islands, that we expect as very dry place. The chosen position is the summit of the island
(320 m). Subsequently, three sites in Sumba island are selected (Sites 2, 7, and 12). Talaud (Site 9) is chosen to represent
northern regions from the equator. In addition, Sangihe-Siau in the extreme north of Indonesia with altitude 1770 m represents
both north of equator and mountain site. 

We define mountain sites as the locations with altitude supposed higher than inversion layer, i.e., higher than $\sim$1000 m
(Varela et al. 2008). Applying our general criteria, there are actually no many suitable choices. We select thus 25 locations: 9 sites at altitudes between 1200 to 2000 m, 10 sites between 2000 to 3000 m, and 6 sites at altitudes higher than 3000 m. Binohoe (Site 23) in Central Sulawesi is located at the closest position to the equator. North of the equator is represented by five sites only.   

We choose four sites in Sumatra at altitudes higher than 2000 m. West Sumatra close to the equator has been surveyed in the 
early 1980s for the plan of giant radio telescope array at metre wavelength (Swarup et al. 1984) which does, in fact, not
really demand excellent atmospheric condition. We select three sites in North Sumatra, near lake Toba, for comparison.

In Java, we select West Ijen (Site 19), located in the far east of Java, as can be suggested from 
clear fraction analysis. This mountain is relatively far from dense population as compared to, for example, Tengger Mountain that is actually a tourist park. In West Nusa Tenggara, in Lombok Island, there is the highest mountain in the region, that is, Mount Rinjani with the summit at altitude 3700 m. We select two positions on this mountain (Sites 25 and 26) that are separated by a distance of about 13 km only, but at significantly different altitudes. The same case is applied to high altitude sites in Papua (Sites 33 and 34) which are separated by a distance of 38 km. East Nusa Tenggara presents ten sites, the most numerous in our selection, including Kupang and Sawu as reference sites. We have visited some locations in
Timor island at the proximity of Sites 3, 8, 29, and 30. Except the latter, our first impression is that the sites are indeed very dry as indicated mainly by the vegetations in the regions. This will be analysed in a separate paper. 

\subsection{Five year mean of clear fraction}

To figure out the general trend of clear fraction for each location, it is useful to consider a yearly mean value
rather than a monthly mean value. Subsequently, we thus compute the yearly mean total clear fraction for each location shown in Table 3 for the last five years (2006--2010). We do not distinguish the total clear fraction between day time and night time as we have seen for monthly mean (Fig.~3 and 4). 

We have mentioned that the year 2010 had shown an extreme weather. At the site of the Bosscha Observatory,
the dry season, which usually occurs in June to September, did not happen. We had so many cloudy days and heavy rainfalls.
We also noted extreme weather reports from various regions in Indonesia that affected badly the transportation and
agriculture. In terms of clear fraction, the value in Lembang dropped to 29 per cent from the average of about 50 per cent of the previous four years. Therefore, the mean clear fraction in Lembang is $\sim$46 per cent. Compared to the sea-level sites, 
it is comparable to Jatiwangi (Site 1) which is in the same region. Clear fraction in Waingapu (Site 2) and Kupang (Site 3) slightly dropped in 2010 as well, but still present clear fraction greater than 65 per cent, and the mean value greater than 70 per cent for the last five years.

Remark that the low-altitude sites (Site 4 to 8) present high clear fraction about 70 per cent, with Sawu (Site 4) gives
the largest (75 per cent). We notice that Sawu presents a value approximately constant for the last five years. It even did not drop in 2010. This is the main reason to take Sawu as reference site.
Sites 4 to 7 are located in small island, and Site 8 in a big island (Timor) but in coastal area.
Their latitudes are below $8^\circ$S. However, this is not the case for Talaud (Site 9) in the extreme north
of Indonesia. It is in a small island too, but gives a clear fraction of 53 per cent only, and at latitude $4^\circ$N.
On the contrary, notice that in 2010 the clear fraction in Talaud increased to 58 per cent. 

Comparison between sea-level and low-altitude sites allows us to retrieve that the best possible clear sky fraction
in Indonesia is about 75 per cent, as indicated by Sawu, and most probably around 70 per cent, as indicated by many locations
in East Nusa Tenggara. Therefore, we expect that potential sites at higher altitudes must be close to these values.
Most regions in Indonesia suffer decreasing value of clear fraction (about 5 per cent) in 2010, but several regions 
keep almost constant value, such as in Sumatra (A) and Sangihe-Siau (Site 24), or even increasing slightly (2--3 per cent), such as in Papua (E).

\begin{figure}
\centerline{\epsfig{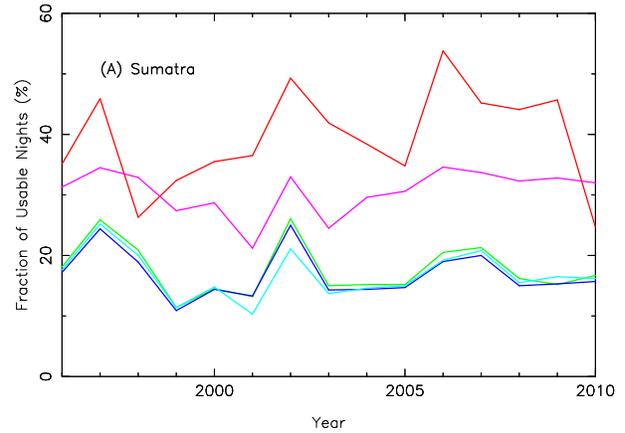}} 
\caption{Yearly usable night fraction, from 1996 to 2010 of the four sites in Sumatra compared to Lembang (Site 10/red).
Only Kerinci (Site 18/magenta) that is slightly comparable to Lembang. Also shown are Site 15 (green), Site 16 (blue),
and Site 17 (cyan).}
\end{figure}
\begin{figure}
\centerline{\epsfig{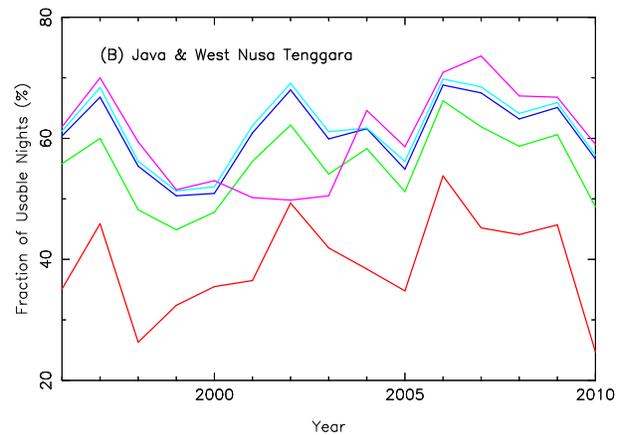}} 
\caption{Yearly usable night fraction, from 1996 to 2010 of the two sites in Java and the three sites in
West Nusa Tenggara. Site 19 (green) in East Java is better than the reference site of Lembang (Site 10/red). The other three sites in West Nusa Tenggara are even better (Site 25/blue, Site 26/cyan, Site 11/magenta). }
\end{figure}
\begin{figure}
\centerline{\epsfig{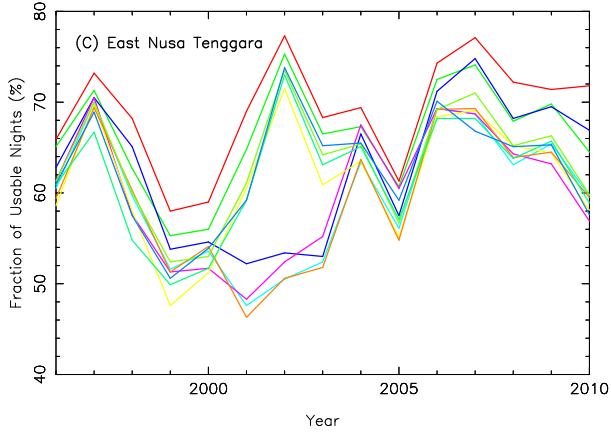}} 
\caption{Yearly usable night fraction, from 1996 to 2010 of the ten sites in East Nusa Tenggara. Kupang (Site 3/green) and Sawu (Site 4/red) are reference sites. The other sites are mostly comparable (Site 27--31).}
\end{figure}

Four high mountain sites in Sulawesi (D) present a clear fraction between 38 to 47 per cent. It means they are not better than the site of the Bosscha Observatory. Except Sangihe-Siau (Site 24) in a small island, it 
provides a higher value of 57 per cent. Moreover, it is comparable to West Ijen in East Java that has $\sim$60 per cent.

\begin{figure}
\centerline{\epsfig{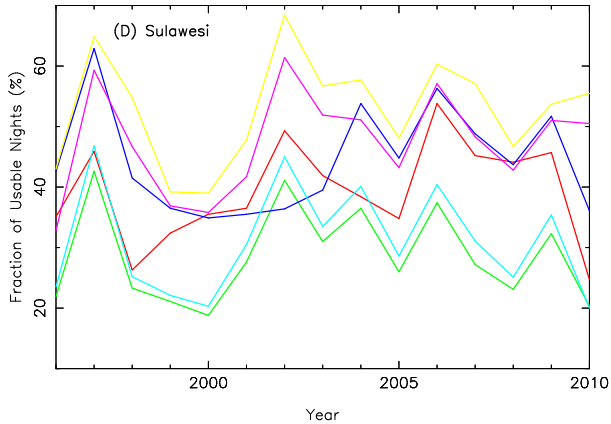}} 
\caption{Yearly usable night fraction, from 1996 to 2010 of the five sites in Sulawesi compared to the reference site of Lembang (Site 10/red). Sangihe-Siau (Site 24/yellow), Binohoe (Site 23/magenta), and Lombosang (Site 21/blue) are slightly better than Lembang.}
\end{figure}
\begin{figure}
\centerline{\epsfig{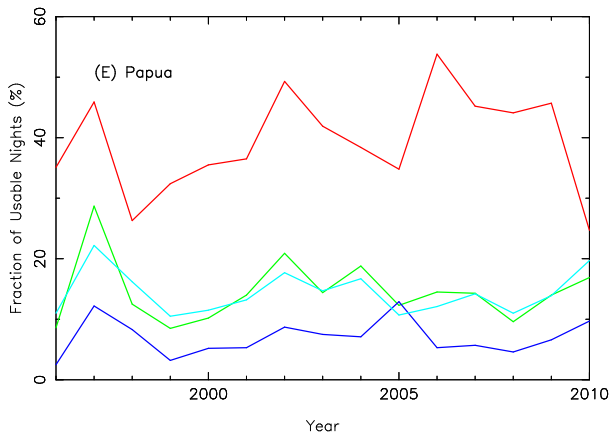}} 
\caption{Yearly usable night fraction, from 1996 to 2010 of the three sites in Papua (Site 32/green, Site 33/blue, Site 34/cyan) compared to the reference site of Lembang (Site 10/red). They are not suitable for astronomical site.}
\end{figure}

The sites in North Sumatra and Papua show low average clear fraction of about 20 per cent. Even the highest summit in Papua (Puncak Jaya 1 at altitude 4700 m) gives a value of 10 per cent only. Hence, they are not promising sites to host a modern astronomical observatory. In the following, however, for confirmation and completeness, these sites are not erased in further analyses. 

\begin{table}																									
\caption{Mean of 15 years of the night fraction in percentage}
\begin{center}																				
\begin{tabular}{lccccc} 													
\hline\hline																									
\mbox{}\hspace{1.cm} Sites	&	Cl. 	&	Tr. 	&	Op. 	&	Usable	&	$\sigma$* \\
\hline
(4) Sawu**	&	65.0	&	4.1	&	30.9	&	69.1	&	5.9	\\
(3) Kupang**	&	61.6	&	4.6	&	33.7	&	66.3	&	6.0	\\
(29) Timau	&	58.6	&	4.6	&	36.7	&	63.3	&	6.3	\\
(12) S. Waingapu 	&	58.2	&	4.5	&	37.3	&	62.7	&	7.8	\\
(31) Alor	&	57.1	&	5.6	&	37.4	&	62.6	&	6.3	\\
(30) Mutis	&	57.1	&	4.7	&	38.2	&	61.8	&	6.5	\\
(26) Rinjani 2	&	56.2	&	5.5	&	38.3	&	61.7	&	6.1	\\
(27) Ruteng	&	56.3	&	5.1	&	38.5	&	61.5	&	6.9	\\
(25) Rinjani 1	&	55.1	&	5.6	&	39.3	&	60.7	&	6.0	\\
(11) E. Sumbawa	&	55.2	&	5.2	&	39.5	&	60.5	&	8.2	\\
(14) Wetar	&	54.3	&	5.6	&	40.1	&	59.9	&	7.3	\\
(13) Sirung	&	54.4	&	5.1	&	40.6	&	59.4	&	7.3	\\
(28) Lembata	&	54.3	&	4.9	&	40.8	&	59.2	&	7.5	\\
(19) West Ijen	&	49.8	&	5.9	&	44.3	&	55.7	&	6.3	\\
(24) Sangihe-Siau	&	44.8	&	8.0	&	47.1	&	52.9	&	8.7	\\
(23) Binohoe	&	39.4	&	8.0	&	52.6	&	47.4	&	8.5	\\
(10) Lembang**	&	33.8	&	5.5	&	60.7	&	39.3	&	8.2	\\
(22) Rantemario	&	25.1	&	6.1	&	68.8	&	31.2	&	8.8	\\
(18) Kerinci	&	24.1	&	6.5	&	69.4	&	30.6	&	3.8	\\
(20) Tibo	&	22.5	&	6.2	&	71.3	&	28.7	&	7.8	\\
(21) Lombosang	&	38.3	&	6.0	&	71.3	&	28.7	&	8.6	\\
(15) Sinabung	&	13.0	&	4.7	&	82.3	&	17.7	&	4.4	\\
(16) Sibayak	&	12.3	&	4.5	&	83.2	&	16.8	&	4.0	\\
(17) Silimapuluh	&	12.2	&	4.5	&	83.2	&	16.8	&	3.9	\\
(32) Erekebo	&	9.3	&	5.2	&	85.5	&	14.5	&	5.3	\\
(34) P. Jaya 2	&	8.9	&	5.4	&	85.6	&	14.4	&	3.5	\\
(33) P. Jaya 1	&	2.9	&	4.0	&	93.0	&	7.0	&	3.0	\\
\hline
\end{tabular}		
\end{center}
\mbox{}\underline{Notes:} *Standard deviation of usable night; **Reference site; Cl. = clear, Tr. = Transitional, Op. = opaque;
the number in the bracket corresponds to the site number (see Table 3).
\end{table}																									

\subsection{Usable nights in five regions}

To obtain the night time clear fraction profile, in the following we analyse mountain sites in terms of yearly average usable night, i.e., a total of clear (photometric) sky and transitional (spectroscopic) sky from 1996 to 2010. 
Fig.~7 shows the results for the sites in Sumatra (A). It appears that from the four selected sites in Sumatra, 
Kerinci (Site 18), altitude 2900 m, is better than the other sites in North Sumatra. However, the largest clear fraction in 
this mountain is low, i.e., $\sim$35 per cent, occured in 1997 and 2006. The other years have lower values with transitional night time contribution of 6.5 per cent. Moreover, the three sites around Lake Toba, at altitudes 2100--2400 m, present much lower value, fluctuating between 10 to 25 per cent. Note that these sites are also close to the equator.

\begin{figure*}
\vbox{
\centerline{\epsfig{figure=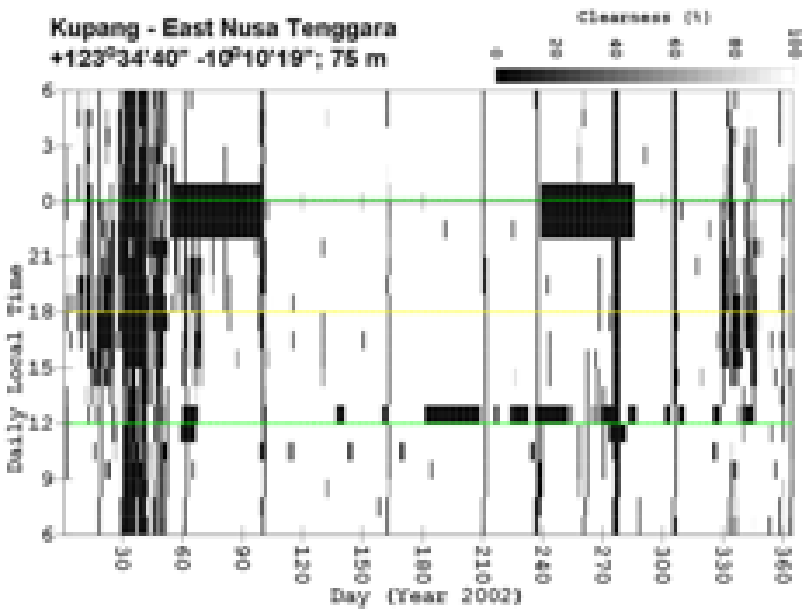,width=7.0cm}\hspace{2mm}\epsfig{figure=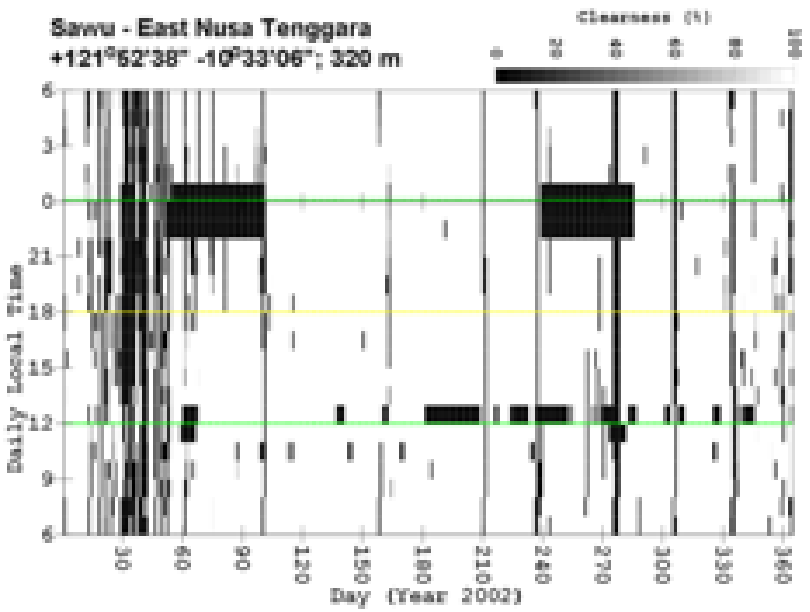,width=7.0cm}}
\vspace{2mm}
\centerline{\epsfig{figure=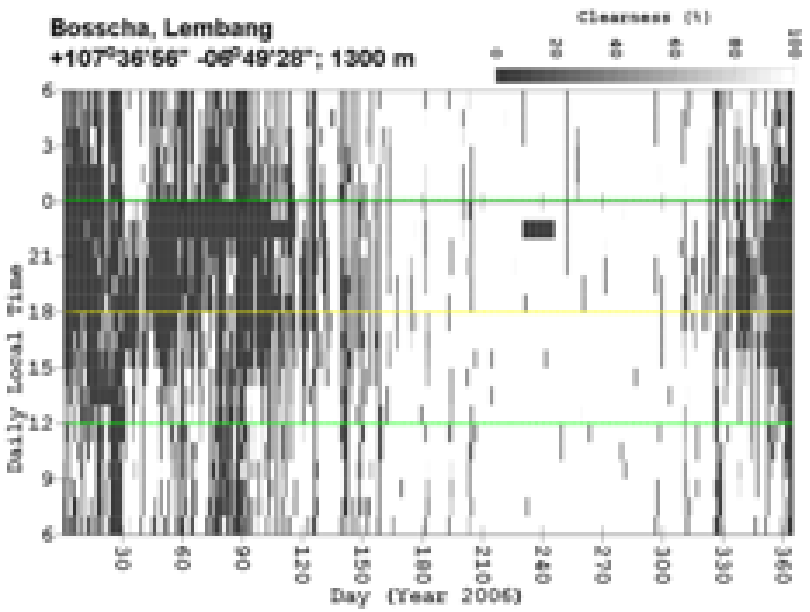,width=7.0cm}\hspace{2mm}\epsfig{figure=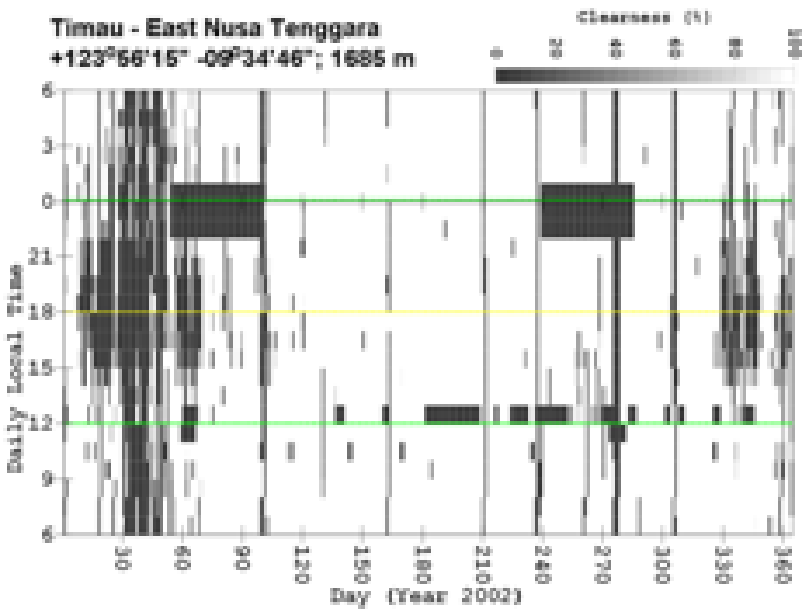,width=7.0cm}}
\caption{Comparison of the daily charts of the reference sites: Kupang (Site 3), 2002 (top-left); Sawu (Site 4), 2002 (top-right);
Lembang (Site 10), 2006 (bottom-left); and Timau (Site 29), 2002 (bottom-right). The indicated year corresponds to the best clear fraction over the past 15 years. Missing data appears as regular black boxes that are considered as opaque pixels.}
}
\end{figure*}
\begin{figure*}
\vbox{
\centerline{\epsfig{figure=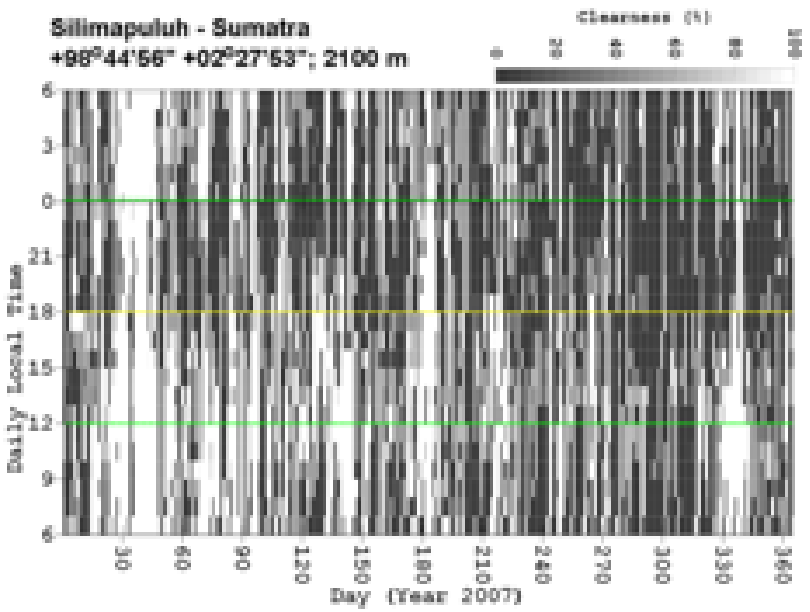,width=7.0cm}\hspace{2mm}\epsfig{figure=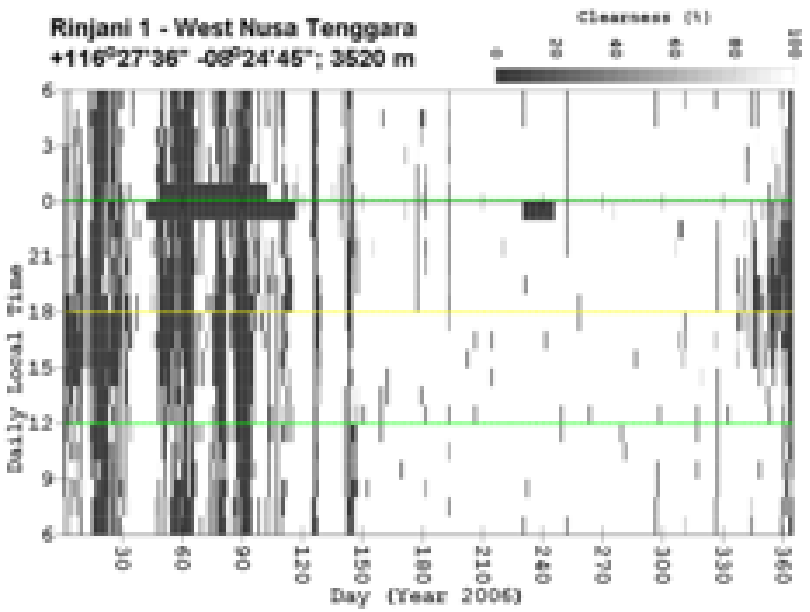,width=7.0cm}}
\vspace{2mm}
\centerline{\epsfig{figure=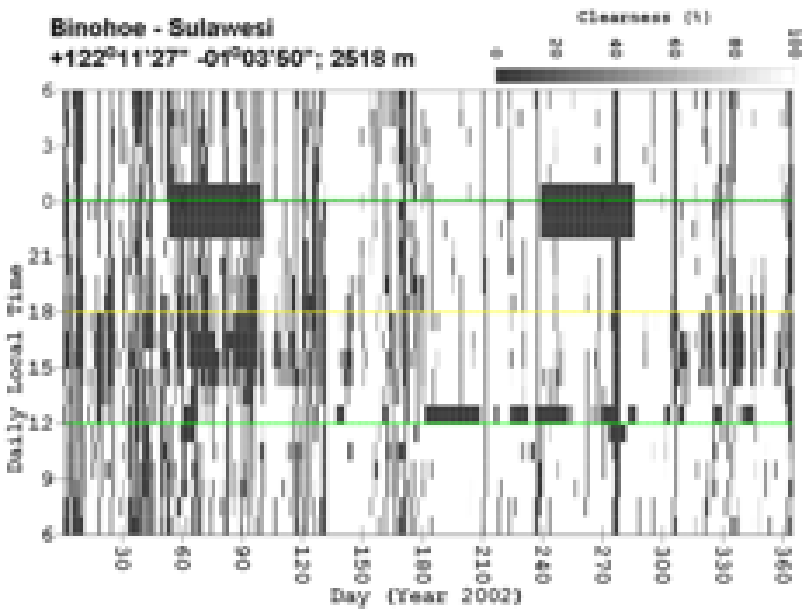,width=7.0cm}\hspace{2mm}\epsfig{figure=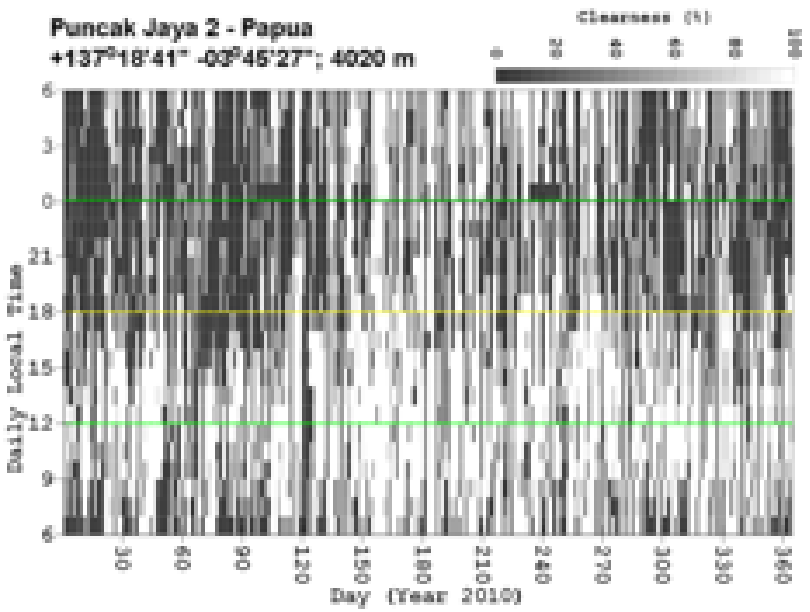,width=7.0cm}}
\caption{Comparison of the daily charts of the sites: Silimapuluh (Site 17), 2007 (top-left); Rinjani 1 (Site 25), 2002 (top-right); Binohoe (Site 23), 2002 (bottom-left); and Puncak Jaya 2 (Site 34), 2010 (bottom-right).}
}
\end{figure*}

Subsequently, Fig.~8 displays comparison of the fraction of clear sky for the sites in Java and West Nusa Tenggara (region B).
First we note that the annual clear night sky in Lembang is about 40 per cent only that corresponds to $\sim$150 nights of
observations (about five months of effective time). Compared to West Ijen (Site 19), we obtain a factor of 
15 per cent better than the reference site of Lembang. High values are found in 1997, 2002, and 2006, i.e., between 60--66 per cent. Spectroscopic nights contribute about 6 per cent for both sites. Furthermore, the sites in Lombok and Sumbawa (Sites 11, 25, 26) present much better clear nights. Since the Sites 25 and 26 are close to each other (despite their different altitudes), we expect that their clear sky fractions are not too different. The largest values ($\sim$70 per cent) are found 
in 1997, 2002, and 2007. East Sumbawa (Site 11), located closer to East Nusa Tenggara rather than Rinjani, appears 
also as a good site but at lower altitudes. 

For region C (East Nusa Tenggara), we have selected two reference sites (Sites 3 and 4) which are found systematically
better than the other mountain sites. Among the eight mountain sites analysed, there are four promising locations, that is,
Timau (Site 29), Alor (Site 31), Mutis (Site 30), and Ruteng (Site 27) with yearly average clear nights above 60 per cent. Spectroscopic nights contribute below 5 per cent, except for Ruteng in Flores Island. The comparison is displayed in Fig.~9.

Moreover, for region D (Sulawesi) and E (Papua), the fraction of clear nights are shown in Fig.~10 and 11, respectively.
They have much lower clear nights compared to C, so they are better compared to the site of Lembang. The mountain sites
of Sulawesi seems providing more fluctuating yearly clear nights. We found only Binohoe (Site 23) and Sangihe-Siau (Site 24)
which are systematically better than Lembang. Note that Lombosang (Site 21) presents a comparable situation to 
Lembang during the last five years. In contrast, the region E (Papua) provides a very low fraction of clear nights (less than
15 per cent). The highest site, Puncak Jaya 1 (Site 33), has clear fraction even mostly below 10 per cent. Therefore, like in North Sumatra, mountains in Papua appear not suitable for astronomical sites.

\subsection{Mean of fifteen years}

Table 4 shows the mean of 15 years of usable nights for the 24 mountain sites including three reference sites. 
The sites are sorted according to decreasing usable nights along with its standard deviation ($\sigma$). We see that our
reference sites 3 and 4 are indeed better with $\sigma\sim 6$ per cent. We also can verify the
contribution of spectroscopic nights in addition to photometric nights. Note that South Waingapu (Site 12), despite its 
large clear fraction, has a higher $\sigma$. The same case holds for East Sumbawa (Site 11) with $\sigma = 8.2$. Therefore, compared to Timau (Site 29), Alor (Site 31), Mutis (Site 29), and Rinjani (Sites 25 and 26),
the Sites 11 and 12 are relatively good but `less stable'. Ruteng (Site 27) also falls in this category. 

We also notice that the contribution of spectroscopic nights are slightly lower in Timau and Mutis compared to
Alor and Rinjani as such that the former are possibly better than the latter. Considering, however, uncertainties on the data ($\sim$10 per cent), we actually estimate that they are likely to be similar. This suggest that we are led to a down-selection to Timor, Alor, and Lombok, three islands in Nusa Tenggara. 

\section{Discussion}

\subsection{Estimation using daily chart}

From the above analysis, we found dry periods, detected in most regions, i.e., in 1997, 2002, and 2006.
So the periodicity of 4--5 years is still to be confirmed in the long period, or decades. Some regions
have persistently show a large clear sky fraction, mainly in East Nusa Tenggara, with its many small islands.
Moreover, coastal area presents better situation of clear sky. Mountain sites in the proximity of coastal area 
generally also present large fraction of the clear sky, such as in Timau and Alor, and also in the north-eastern side of
Rinjani. 

The yearly clear sky pattern can be inspected quickly using a daily chart as can be seen in Fig.~12 and 13.
In Fig.~12, we display three charts of reference sites (Kupang, Sawu, and Lembang) as compared to Timau, a potential
site inferred in this study. The chart of Lembang is for 2006, that is, the driest period of the 15 years, while 
the three others are for 2002, corresponding to the same driest situation. Notice also missing data (appearing like
regular black boxes) that we consider in our calculation as opaque. We see that cloudy situations are found mostly in January
and February in the three sites of Nusa Tenggara, compared to longer cloudy situation in Lembang, from January to April,
and starting over again in December of the corresponding year. 

Then, we also select the driest period for four other less clear sites as shown in Fig.~13, i.e., the sites of
Silimapuluh (in A), Rinjani 1 (in B), Binohoe (in D), and Puncak Jaya 2 (in E). We have a clear sky in Silimapuluh
only between January and February and then almost opaque all the time. The case of Rinjani is similar to the sites presented in Fig.~12, but it is slightly less clear than that in East Nusa Tenggara. For Binohoe, the situation has a smaller fraction of clear sky. Lastly, for Puncak Jaya, it appears that the cloud in this region presents almost every time and the night time appears more cloudy than the day time, probably due to biases introduced by the fixed thresholds.   

\subsection{Potential sites}

As inferred in Section 5.4, Timau in Timor island is relatively the best site in terms of the fraction of usable night.
However, the altitude of this site is lower than 2000 m. Mutis, about 45 km to the south-east of Timau, is also good
site with a high altitude of 2400 m. Timau is relatively better than Alor (about the same altitude) because Timau has
more photometric nights. We have visited both Mutis (in September 2007) and Timau (in August 2008 and 2009), but not up to the summits. At the time of our visit to Mutis, we found it was wet, and there were fog at altitudes lower than 1600 m. On the other hand, the area around Timau is indeed very dry.  

The site of Rinjani in Lombok island is also relatively good, and we prefer the point to the north-east of the
mountain summit. The photometric night fraction is 55 per cent while the spectroscopic fraction is slightly greater than 5 per cent. Ruteng in Flores island cannot be ruled out from this analysis since it is marginally comparable to Rinjani.
We recommend therefore four Island in Nusa Tenggara for further site testing, with the sites in Timor island as high priority. 

The present work is limited to identifying the potential sites for our future national observatory.
More detailed analyses for these sites in a monthly average fraction will be presented in a forthcoming paper.

\section{Conclusions}

Meteorological satellite data with a time span of 15 years, from four different satellites, have been used to retrieve
the fraction of clear sky above Indonesia for astronomical site selection. This allows us 
to obtain a detailed cloud cover pattern over Indonesia since the data have high spatial resolution ($\sim$5 km) 
and provide a time span longer than a decade. Accordingly, the fraction of clear sky can be analysed separately
for day time and night time, and either for monthly or yearly average.
The fraction of photometric and spectroscopic nights are inferred to obtain the usable nights, and also the fraction
of opaque nights. Fifteen years data period has indicated that clear sky fraction over some regions in East Nusa Tenggara can 
unambiguously reach better than 70 per cent with an uncertainty of 10 per cent. A stable clear night fraction is found during April to October with a value between 55 to 75 per cent. 

In tropical region like Indonesia, we find that the higher the altitudes of the regions (mountains), the more frequent the presence of clouds is. Coastal regions in East Nusa Tenggara are systematically drier than that in the high mountains.
Five sites are recommended for site testing and to be analysed further: Timau, Mutis, Alor, Rinjani, and Ruteng. 
Regions near Timau in Timor has been visited, but a longer and complete survey must also be conducted.
Site survey is absolutely necessary to obtain real situation in the field and we must undertake in situ measurement continuously. This also must consider the detailed geographic situation of the regions.  
A number of theoretical and measurement programmes will be undertaken in the near future. 

\section*{Acknowledgments}

The authors wish to thanks the weather team in Kochi University, Japan, for providing the data archive open for research and education. The data are available at Weather Image Archive, Kochi University, downloadable at http://weather.is.kochi-u.ac.jp/archive-e.html. We thank very much the reviewer, Dr. S. Ortolani, for his critical and careful reading which improve this paper. This work is supported by KNRT--Indonesia Research Grant, through {\it Program
Insentif Ristek} 2007--2009, to whom we are sincerely indebted.


\bsp

\label{lastpage}

\end{document}